\RequirePackage{fix-cm}
\documentclass[twocolumn]{svjour3}          % twocolumn
\smartqed  % flush right qed marks, e.g. at end of proof
\usepackage{graphicx}
\usepackage{amsmath}
 \usepackage{amsfonts}
\usepackage{algorithmicx,algorithm}
\usepackage{cite}
\begin{document}

\title{Concurrent Encryption and Authentication for Wireless Networks using Compressed Sensing%\thanks{Grants or other notes
%about the article that should go on the front page should be
%placed here. General acknowledgments should be placed at the end of the article.}
}
%\subtitle{Do you have a subtitle?\\ If so, write it here}

%\titlerunning{Short form of title}        % if too long for running head

\author{Chaoqing Tang
}

%\authorrunning{Short form of author list} % if too long for running head

\institute{Chaoqing Tang \at
              School of Artificial Intelligence and Automation, Huazhong University of Science \& Technology, Wuhan 430074, China. \\
              \email{billtang@hust.edu.cn}             
%             \emph{Present address:} of F. Author  %  if needed
}     

\date{Received: date / Accepted: date}
% The correct dates will be entered by the editor

\maketitle

\begin{abstract}
Authentication and encryption are traditionally treated as two separate processes in wireless networks, this paper integrates user authentication into the process of solving eavesdropping attacks. A compressed sensing (CS)-based framework is proposed which manipulates the measurement matrix of CS to safeguard secure computationally. The framework is also capable of continuous authentication and transmission error correction and is robust to data loss. In detail, this paper first proposes an algorithm to generate a 2D key which depends on the physical property of communication channels. The 2D key is further used to generate authentication information and signal structure as well as encrypt original data. Then an encrypted message which contains both data and authentication information is formed for anonymous transmission. The legal receiver can split authentication information and data, and performing a data loss-robust and transmission error-robust authentication and recovery strategy. The framework is evaluated quantitatively using Monte Carlo simulation with simulated sparse signal. The secure transmission performance and authentication performance as well as data loss robustness are investigated. This framework provides an integrated security solution that efficiently safeguards the confidential, privacy and robust communication in cyber-physical systems, especially in resource-limited and safety-critical wireless networks.
% \PACS{PACS code1 \and PACS code2 \and more}
% \subclass{MSC code1 \and MSC code2 \and more}
\end{abstract}

%====================================================
\section{Introduction}\label{sec1}
Modern Internet is innovating and inventing its way to pervasive always-connected broadband services. Aiming to offer cyber-physical systems (CPS) which connect the entire world seamlessly and ubiquitously between anybody, anything, anywhere, anytime, by whatever electronic devices/services/networks \cite{RN522}, which is the Internet of Things (IoT) vision \cite{RN523}. The pervasive data gathering arouses big data issues. In dealing with privacy and security concerns, authentication and encryption are two enabling technologies \cite{RN524}. Authentication is the process of verifying the user's identity, which is the prerequisite to authorization. Encryption is the process of encoding information so that only authorized parties can access it. Traditional authentication and encryption techniques need complex key manage scheme which cannot fit with the resource-limited networks in IoT, e.g. wireless structural health monitoring networks \cite{RN564}. Physical layer security is promising to address these concerns by exploring the characteristic of physical attributes in cyber-physical systems. Compared to traditional key-based authentication and encryption, the specific physical-layer attributes are directly related to the communicating devices and the corresponding environment, which are extremely difficult to impersonate. Some other issues like data loss are crucial as well. Data loss usually happens in wireless sensor networks due to ill quality of wireless channel, unexpected failure, network congestion etc., which harms data integrity. Given that IoT integrates heterogeneous networks, a review on techniques to handle these challenges in cellular networks, wireless sensor network, wireless body area networks etc. are given in the next subsection.
%-------------------------------------------------------

\subsection{Related Works}\label{sec1.1}

Physical layer security explores the characters of communication entities to enable confidential communication in the presence of unauthorized eavesdroppers. A. Mukherjee et al. \cite{RN537} conducted a survey on the principles of physical layer security in multiuser wireless networks. They categorize the enable technologies into a group that dispensing with a secret key by intelligent transmit coding designing \cite{RN527}, and a group that exploiting the wireless medium to generate secret keys over public channels. A review targeting at the wireless physical layer authentication is given by \cite{RN529}. Authors divide existing physical layer authentication schemes into channel-based and analog front end (AFE)-based. Channel-based technologies make use of channel features like the received signal strength indicator (RSSI) \cite{RN535} and channel state information (CSI) including channel gain \cite{RN533, RN526}, multi-path delay \cite{RN530}, carrier phase \cite{RN534, RN532, RN528}, signal power spectrum density \cite{RN536} etc. These features are usually location-sensitive. A distance between adversary and legitimate user that is greater than the half of the wavelength leads to different CSI \cite{RN530}, which is robust to spoofing attack. AFE-based techniques explore some device-specific characteristics, which also referred to as fingerprint \cite{RN525, RN531}. Examples including components character such as the power amplifier and digital-to-analog converter \cite{RN539}, in-phase/quadrature imbalance \cite{RN538}, and the carrier frequency offset \cite{RN540}. Compared to channel based methods, the hardware attributes of transmitter need to be known by receiver in advance.
In terms of physical layer encryption, all the above-mentioned attributes can be used to generate encryption keys \cite{RN537}. T. R. Dean et al. \cite{RN549} explore CSI in massive multiple-input-multiple-output (MIMO) systems to generate CSI-based key. C. Liu et al. \cite{RN552} propose a location-based encryption framework. The location can be estimated by received signal strength (RSS), angle of arrival (AOA), time of arrival (TOA) etc. M. Wilhelm et al. \cite{RN553} use frequency-selectivity of multi-path fading channels for key generation. A study on the sub-carrier obfuscation is carried out under orthogonal frequency-division multiplexing (OFDM) scheme in \cite{RN544}. Investigation on visible light communication channel for key generation also attract some attentions \cite{RN550}. Besides exploring physical attributes for key generation, smart coding strategies also makes an important part of physical layer encryption. These coding include polar code \cite{RN541} and precoding matrix before MIMO transmission \cite{RN551} etc. There are also other initiatives on using interference \cite{RN543} or noise \cite{RN546} for physical layer encryption.

All these proposals contribute to the thriving of solo authentication or encryption on physical layer. This paper aims to take a further leap which achieve concurrent encryption and authentication by using compressed sensing (CS). E. Cand\'{e}s, J. Romberg, and T. Tao ignite the CS in 2006 by demonstrating the rationale of CS theory from the mathematical perspective \cite{RN4}. Compared to Shanon-Nyquist sampling theorem, CS able to reconstruct original signals from far fewer samples than that of Nyquist sampling rate by finding sparse solutions to underdetermined linear systems. Sparsity, which is a prerequisite for CS, means only a small percentage of components in a signal are non-zeros. Since then, CS has witnessed an explosive attention in many fields \cite{RN25,RN460}. 

CS also witnesses some applications in security enhancement. The widely adopted method is using the measurement matrix in compressed sensing as an encryption key \cite{RN457, RN514,RN766}, because only receivers who know the measurement matrix can reconstruct the original signal successfully. It is demonstrated that CS achieves perfect secrecy under conditions if the measurement matrix holds the restricted isometry property (RIP) and the measurement results are more than two times of the sparsity level while the message set does not include the zero message \cite{RN558}. Some literatures \cite{RN556, RN555} also add artificial noise in CS-based encryption method in cooperative radio systems. All these literatures are dealing with physical layer encryption only. CS also shows potential in dealing with data loss problem in wireless sensor networks \cite{RN559, RN38,RN767} due to the downsampling ability of it.
%--------------------------------------------------
\subsection{Contributions \& Paper Structure}
Compared to all the related works which regard encryption and authentication as separate processes, this paper proposes a CS-based framework to perform encryption and authentication concurrently from physical layer perspective. Under this framework, the traditional authentication before access to data mode is simplified by concurrent authentication and encryption/secure transmission, thus providing solution to deal with security and privacy simultaneously. The framework eliminates complex key management which means suitable for resource-limited devices, and is capable of continuous authentication because authentication information goes parallel with data. Besides, this framework has a certain ability of transmission error correction and is robust to data loss. In more detail, this paper first proposes an algorithm to generate a 2D key using channel gain of time-varying channels. The 2D key working as measurement matrix is further used to generate authentication information and signal structure as well as encrypt data by compressed sensing measurement. Then an encrypted message which contains both data and authentication information is formed for secure transmission. The legal receiver can extract authentication information and perform a data loss-robust and transmission error-robust authentication and recovery strategy. Although the measurement matrix is generated using the channel gain, it is easy to transfer to other physical attribute of CPS.

The rest of this paper is organized as follows. Section II describes the preliminaries including system model and compressed sensing theory. Section III introduces the proposed schemes. The performance evaluations are given in Section IV before the conclusions in Section V.
%===========================================
\section{PRELIMINARIES}
\subsection{System Model}
\begin{figure}[!ht]
	\centering
	\includegraphics[width=8.5cm]{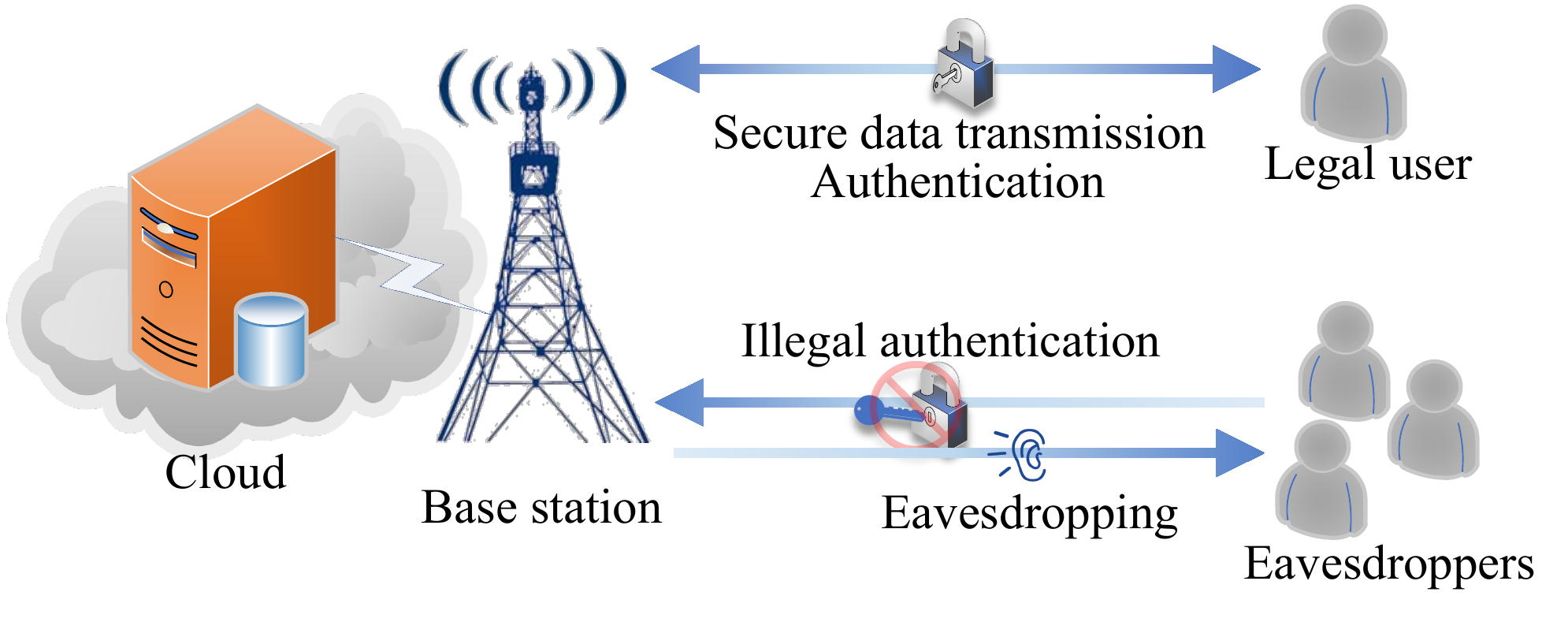}
	%\vspace{-0.5cm}
	\caption{Typical wireless scene for cyber-physical systems. The system suffers from eavesdropping attack and spoofing attack, which need encryption and authentication.}
	\label{fig1}
\end{figure}

The front-end of IoT cyber physical systems provide compatibility with multiple wireless standards, e.g. Zigbee, cellular network, WiFi. The devices in same wireless coverage communicate with the cloud through base station or gateway, such as the 6LowPAN standard. This paper considers a wiretap scene in a wireless network as shown in Fig. \ref{fig1}. A legal user communicates with the base station with certain secure data transmission schemes and both the base station and the legal user need to authenticate each other. Meanwhile, some eavesdroppers attempt to be authenticated by the base station and eavesdrop the downlink to the legal user. Traditional methods use pre-shared key to encrypt transmission data for eavesdropping attack, and use some digital identity sequence for authentication and spoofing attack. This paper investigates physical layer attributes to generate key which eliminates the demand of pre-shared keys and complex key management.
%\vspace{-0.5cm}

We assume all wireless channels in this scene experience Rayleigh fading. Rayleigh fading is applicable for scenes like heavily built-up city centres where there are no line of sight between the transmitter and receiver and many buildings and other objects attenuate, reflect, refract, and diffract the signal. In this case, the channel fading amplitude   is distributed according to
\begin{equation}
p\left( h \right) = \frac{{2h}}{\Omega }\exp \left( { - \frac{{{h^2}}}{\Omega }} \right),{\rm{ }}h \ge 0\
\label{eq1}
\end{equation}
where $\Omega$ is the mean square value of $h$. Note that this is not a spotlight for this paper, one can easily transfer to other fading channels. As a core assumption of this paper, legal user and eavesdroppers experience different channel conditions, this is practical and common if they located far enough apart. All users are assumed to employ multiple-input-multiple-output (MIMO) techniques with $N$ channels and maximal-ratio combining (MRC), this is a general assumption which includes $N = 1$ for single antenna devices. The received symbol signal can be denoted as \cite{RN561},
\begin{equation}
y = x\sum\limits_{i = 1}^N {{h_i}^2\sqrt {{E_{i}}} }  + \sum\limits_{i = 1}^N {{h_i}{w_i}}
\label{eq2}
\end{equation}
where ${E_{i}}$ and ${w_i}$ are the symbol amplitude and Gaussian noise for individual channels, respectively.
%----------------------------------------------------------------
\subsection{Compressed Sensing}
Compressed sensing enables sampling bellow Shanon-Nyquist sampling rate based on finding sparse representation to original signals. The overall principle of CS is stated here.

If a signal ${\mathbf{x}} \in {\mathbb{R}^{n \times 1}}$ can be expanded on a basis  as ${\mathbf{\Psi }} \in {\mathbb{R}^{n \times n}}$ as ${\mathbf{x}} = {\mathbf{\Psi x'}}$, and  has only  $K$ non-zero values (where $K \ll n$), the signal can be measured using a measurement matrix ${\mathbf{\Phi }} \in {\mathbb{R}^{m \times n}}$ (where  $m \ll n$) as
\begin{equation}
{\mathbf{y = {\Phi}x + \xi = {\Phi}{\Psi}x' + \xi = Ax' + \xi }}
\label{eq3}
\end{equation}
where ${\mathbf{y}} \in {\mathbb{R}^{m \times 1}}$ is the measurement results; ${\mathbf{A}} \in {\mathbb{R}^{m \times n}}$  and the $i$-th column of ${\mathbf{A}}$ is denoted as ${{\mathbf{A}}_i}$; ${\mathbf{\xi }} \in {\mathbb{R}^{m \times 1}}$ is Gaussian noise. It is worth noting that the dimension of measurement results is much smaller than that of the original signal since $m \ll n$, which means the original signal is compressed when it is sampled.

Recovering ${\mathbf{x'}}$ using ${\mathbf{y}}$ and ${\mathbf{A}}$ is to solve the following optimization problem:
\begin{equation}
\mathop {\min }\limits_{{\mathbf{x'}}} {\text{  }}{\left\| {{\mathbf{x'}}} \right\|_{\ell _0}}{\text{ subject to }}{\left\| {{\mathbf{y}} - {\mathbf{Ax'}}} \right\|_{\ell _2}} \leq \tau
\label{eq4}
\end{equation}
where ${\left\|  \cdot  \right\|_{\ell _0}}$ and $\left\|  \cdot  \right\|_{\ell _2}$ denote  ${\ell _0}$-norm and  ${\ell _2}$-norm, respectively. $\tau$ is a tolerance. Because (\ref{eq4}) is a NP-hard problem. It is widely accepted to substitute  ${\ell _0}$-norm using the closest  ${\ell _1}$-norm under some conditions like the RIP \cite{RN272} or mutual incoherence property (MIP) \cite{RN310}. In industrial process, MIP is preferred option due to lower computation complexity than RIP \cite{RN88}, it is defines as,
\begin{equation}
\mu  = \mathop {\max }\limits_{i \ne j} \left| {\left\langle {{{\mathbf{A}}_i},} \right.\left. {{{\mathbf{A}}_j}} \right\rangle } \right| \label{eq5}
\end{equation}
where $\left\langle { \cdot ,} \right.\left.  \cdot  \right\rangle$ is the inner product and $\left|  \cdot  \right|$ is the absolute value function. E. J. Cand\'{e}s et al. \cite{RN272} proved that 0/1-Bernoullic matrices and independent identically distributed Gaussian matrices guarantee sparse recovery in most cases under measurement number constraint,
\begin{equation}
\alpha {\mu ^2}K\log n \le m \le n     \label{eq6}
\end{equation}
where $\alpha$ is a factor corresponding to different instance. This constrain means that smaller mutual incoherence leads to fewer samples requirements. These conditions guarantee high probability recovery using variety of recovery algorithms like orthogonal matching pursuit (OMP) \cite{RN18}. 

Finally, Combining the recovered sparse signal with the sparse basis to get ${\bf{x}}$ as ${\bf{x = \Psi x'}}$.

\section{PROPOSED SCHEMES}
This section introduces the proposed schemes. An overall diagram is given following by each block details.
%-----------------------------------------
\subsection{Overall framework}
\begin{figure}[!ht]
	\centering
	\includegraphics[width=8.5cm]{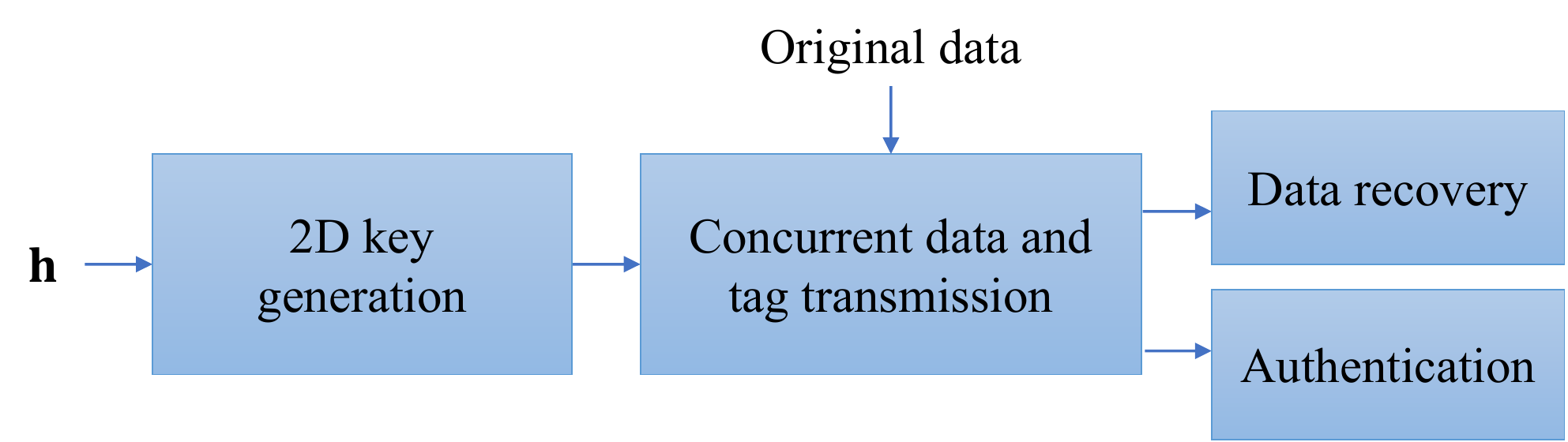}
	\caption{Overall diagram of the proposed concurrent encryption and authentication method.}
	\label{fig2}
\end{figure}

Fig. \ref{fig2} depicts the overall diagram. Firstly, the channel gains are used to generate a 2D key (measurement matrix) through a key generation process. These channel gains can be estimated from individual MIMO channels. Secondly, the measurement matrix samples the signal and generate authentication information (which referred to as tag) in the concurrent data and tag transmission block. This block is also responsible for hiding tag in the measurement results of original data. Finally, the receiver splits the tag and data, and reconstructs the original data and authenticates the transmitter.

It is worth noting that this is a single way transmission, only the receiver can authenticate the transmitter. In our scene, the eavesdroppers attempt to get authentication by the base station, thus the transmitters are eavesdropper and legal user for the uplinks. The base station responses to legal receiver in the downlink to get reverse authentication. Under this scheme, all participants send confidential information while reserving the authentication ability. Authentication and encryption information are parallelly transmitted. This feature provides low latency communication, because authentication and encryption are serial process in data access traditionally.
%----------------------------------------------------
\subsection{2D key generation}
As discussed in section II.B, independent identically distributed Gaussian matrices are suitable for measurement matrix in most cases. R. Dautov et al. \cite{RN457} provides a Gaussian and Bernoullic matrices generation methods based on channel gain and m-sequence. This method is adopted in \cite{RN35, RN514}. But the method requires a very long channel gain sequences, which brings in long communication latency. Compare to \cite{RN457}, instead of requiring a very long channel gain sequence, this paper relaxes it greatly by using a random cyclical shift to a single m-sequence, the shift value is decided by the short group of channel gain. In this way, both transmitter and receiver saves a lot of time in channel estimation before transmitting and authentication.
\begin{algorithm}[!ht]
	\caption{Measurement matrix generation}
	\hspace*{0.02in} {\bf Input:} Channel gain, $\mathbf{h} = \left \{ h_1, h_2, h_3, ..., h_L \right \}$.
	\begin{algorithmic}[1]
		\State $seed =\varepsilon \left ( \mathbf {h}-E\left [ \mathbf{h} \right ] \right ) $ , where $\varepsilon \left ( \cdot \right )$ is the unit step function.
		\State Get $L-1$ order power primitive polynomial.
		\State Use the $seed$ as initial value of linear feedback shift register (LFSR) and the primitive polynomial as feedback function to get m-sequence.
		\State Add a 0 to the m-sequence to balance the number of 0 and 1, denote as $\mathbf{m}$, the length is $2^{L}$.
		\State Cyclically shift the new m-sequence according to $\mathbf{h}$ and store every results, $\mathbf{m}$. A simple shift sequence is rounding $f_s(\mathbf{h}) = (2^L \sum_c \mathbf{h})/\sum \mathbf{h}$, where $\sum_c$ is cumulative sum. The shift number is $L$ in this case. One can flip $\mathbf h$  or even use the summation of original $\mathbf h$  and flipped $\mathbf h$  as the input for $f_s(\cdot)$  to obtain a longer shift sequence.
		\State	Sum all $\mathbf{m}$ to get a $2^L$  sequence.
		\State Normalize to 0$\sim1$ and reshape into a matrix ${\mathbf{\Phi }} \in {\mathbb{R}^{m \times n}}$.
	\end{algorithmic}
	\hspace*{0.02in} {\bf Return:} measurement matrix $\mathbf{\Phi}$.
\end{algorithm}

A detailed description of the proposed method is given in Algorithm 1. Firstly, $L$ channel gains are used to generate a seed which consist of only binary values. Then a $L-1$ order power primitive polynomial is obtained by consulting look-up tables or calculation. Followed by a m-sequence (a basic pseudo-noise sequence in CDMA) generation process which realized by linear feedback shift register (LFSR). A `0' is concatenated to the m-sequence to balance the number of `0' and `1' and also to make up an even number sequence. Different from \cite{RN457} which repeatedly estimates channel gains and repeats step (1) to (4) for many times to get different m-sequences, this paper proposes a cyclical shift scheme as shown in step (5). The cyclical shift scheme is way faster than repeatedly obtaining new channel gains, thus reducing latency. After the following steps in Algorithm 1, a Gaussian matrix is obtained which works as 2D key for encryption and the measurement matrix for CS.

\begin{figure}[!ht]
	\centering
	\includegraphics[width=8.5cm]{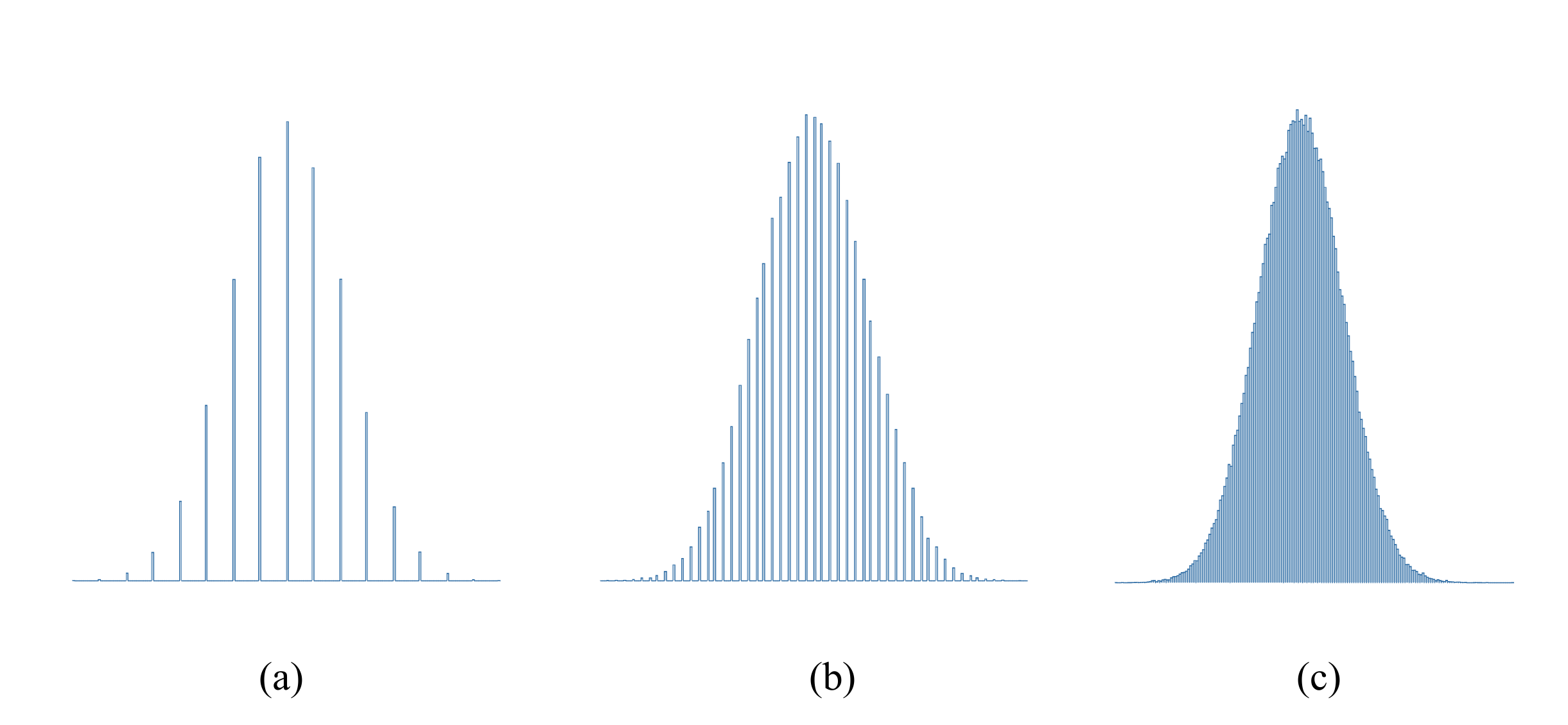}
	\caption{The influence of cyclical shift times on measurement matrix. (a) distribution after shifting $L$ = 18 times; (b) distribution after shifting $6L$ times; (c) true Gaussian distribution.}
	\label{fig3}
\end{figure}

The cyclical shift times influence the accuracy of generated Gaussian matrix. More shift times lead to more accurate results. The generated Gaussian measurement matrix is discrete as shown in Fig.\ref{fig3}. With the increasing of shift times, the discreteness becomes less obvious. Actually, the true Gaussian distribution is discrete because those continuous values are limited by the storage precision of digital systems. From this perspective, the proposed measurement matrix generation process can realize true Gaussian distributed matrix under enough shift times.
%--------------------------------------------------------------------------
\subsection{Concurrent data and tag transmission}
The traditional CS-based encryption achieves secure transmission by transmitting the sampled results rather than the original data. The measurement matrix works as a key to the original data. This paper integrates this scenario with authentication ability.

Considering an original signal ${\bf{x}} \in {\mathbb{R}^{n \times 1}}$ samples by the generated measurement matrix ${\bf{\Phi }} \in {\mathbb{R}^{m \times n}}$ as,
\begin{equation}
{\bf{y}} = {\bf{\Phi x + \xi }}
\label{eq7}
\end{equation}
where ${\bf{y}} \in {\mathbb{R}^{m \times 1}}$ is the weighted sum of each row in ${\bf{\Phi }}$ in fact, the weighting factor is ${\bf{x}}$, which can be rewritten as
\begin{equation}
{y_i} = \sum {{\Phi _{ij}}{x_j}}  + {\xi _i}
\label{eq8}
\end{equation}
where $i = \left\{ {1,2,3,...,m} \right\}$, $j = \left\{ {1,2,3,...,n} \right\}$. Then we propose a concurrent data and tag transmission scheme. Firstly, a tag index which only consist of `0' and `1' is introduced,
\begin{equation}
{k_i} = \left\{ {\begin{array}{*{20}{c}}
	{1,}&{{\rm{if }}\prod\limits_{j = 1}^n {\left( {{\Phi _{ij}} - 1} \right) = 0} }\\
	{0,}&{{\rm{otherwise}}}
	\end{array}} \right.
\label{eq9}
\end{equation}
As the measurement matrix is normalized between 0 and 1, `1' is the maximal value in ${\bf{\Phi }}$. (\ref{eq9}) is finding rows where the maximal values appear. Then a tag sequence is generated according to (\ref{eq10})
\begin{equation}
{t_i} = {k_i}{\Phi _{ii}}
\label{eq10}
\end{equation}
Some elements in ${\bf{\Phi }}$ are chosen as the tag value. It is obvious that the tag only related to the measurement matrix, which related to the channel gains ultimately. This authentication information generation process in (\ref{eq9}) and (\ref{eq10}) do not have complex computation task such as Hash function, which is easy to implement by either hardware or software. Finally, substituting the tag index position in measurement results as tag to obtain the transmitting message as
\begin{equation}
{s_i} = {y_i}\left( {1 - {k_i}} \right) + {k_i}\left( {{t_i}{\rm{ + }}{{\sum\limits_{p = 1}^m {{y_p}\left( {1 - {k_p}} \right)} } \mathord{\left/
			{\vphantom {{\sum\limits_{p = 1}^m {{y_p}\left( {1 - {k_p}} \right)} } {\sum\limits_{p = 1}^m {\left( {1 - {k_p}} \right)} }}} \right.
			\kern-\nulldelimiterspace} {\sum\limits_{p = 1}^m {\left( {1 - {k_p}} \right)} }}} \right)
\label{eq11}
\end{equation}
Besides substituting, this process adjusts the amplitude of the tag sequence to the same level of the measurement results, which makes eavesdroppers hard to distinguish tag positions. More robustly, the tag number is variable. According to the Gaussian distribution, the probability of maximal value is low, which means only a small part of measurement results will be substituted as tag.

After (\ref{eq9}) to (\ref{eq11}), the tag is hiding into the transmission message. A different measurement matrix leads to different substituting position, i.e., different signal structure. The legal user can reconstruct the original data perfectly even only part of the measurement results are transmitted which will be discussed in the next subsection. This means that concurrent data and authentication information transmission is achieved.
%------------------------------------------------------------------------
\subsection{Data recovery}
As the legal user and the base station share the same wireless channel, the receiver can obtain the same measurement matrix as the transmitter, and obtain the same tag index $k_i$ as a result. The received measurement results can be extracted easily using tag index. Unfortunately, there may be some transmitting error in the received measurement results. Those error symbols usually show up as extreme values, some filtering process can be introduced to get rid of these wrong values. Sometimes even completely loss data in some time slots. Regardless of the reason behind data loss, the final received measurement results (denote as ${\bf{\hat y}}$ ) is a subset (denote the subscript set as $\bf{i}$ of ground truth. (\ref{eq8}) tell us that every single measurement result value contains the overall information of original signal in some degree. The constrain for perfect recovery also gives in (\ref{eq6}), so the original information can be reconstructed perfectly by using ${\bf{\hat y}}$, ${{\bf{\Phi }}_{\bf{i}}}$ and the corresponding sparse basis ${\bf{\Psi }}$ as
\begin{equation}
\mathop {\min }\limits_{{\bf{\hat x}}} {\rm{  }}{\left\| {{\bf{\hat x}}} \right\|_{{\ell _0}}}{\rm{\text{ }subject\text{ }to\text{ }}}{\left\| {{\bf{\hat y}} - {{\bf{\Phi }}_{\bf{i}}}{\bf{\Psi \hat x}}} \right\|_{{\ell _2}}} \le \tau
\label{eq12}
\end{equation}

The transmission errors in the physical layer are corrected, which indicates the error-correction ability.
%---------------------------------------------------------------------------
\subsection{Authentication}
The legal receiver can extract the tag part using the tag index and extract the tag $\hat t$ as
\begin{equation}
\begin{array}{l}
{\bf{\hat t}} = \left\langle {{\bf{\hat s}},{\bf{k}}} \right\rangle  - {{\sum\limits_{p = 1}^m {{y_p}\left( {1 - {k_p}} \right)} } \mathord{\left/
		{\vphantom {{\sum\limits_{p = 1}^m {{y_p}\left( {1 - {k_p}} \right)} } {\sum\limits_{p = 1}^m {\left( {1 - {k_p}} \right)} }}} \right.
		\kern-\nulldelimiterspace} {\sum\limits_{p = 1}^m {\left( {1 - {k_p}} \right)} }}\\
{\rm{   = }}\left\langle {{\bf{\hat s}},{\bf{k}}} \right\rangle  - E\left[ {{\bf{\hat y}}} \right]
\end{array}
\label{eq13}
\end{equation}
where ${\bf{\hat s}}$ is the received message, $\bf k$ is the tag index set. Under low data loss ratio and transmission error, the mean value of the original measurement results part in the transmitting message is same as that of the final received measurement results. To authenticate the transmitter, the receiver only need to compare the received tag and the generated tag. Because the tag only consists of a small part of the transmitting message, it is too sensitive to transmission errors and data loss if use root-mean-square error (RMSE) as cost function. We relax it by comparing tags one-by-one and allow their difference below certain threshold. If the identical tag number is greater than a threshold, it is classified as authenticated. The process is represented mathematically in (\ref{eq14}),
\begin{equation}
\sum\limits_{q = 1}^{{N_t}} {\varepsilon \left( {{\tau _1} - \left| {{{\hat t}_q} - {t_q}} \right|} \right)}  > {\tau _2}{N_t}
\label{eq14}
\end{equation}
where ${N_t}$ is the tag number, $\tau_1$ and $\tau_2$ are two thresholds. Under such setting, the authentication is more robust to the data loss and transmission error. The eavesdropper cannot know either the signal structure/tag position nor the tag numbers, not to mention tag values.
%=======================================================================
\section{PERFORMANCE EVALUATION}
The framework is evaluated in terms of secure transmission performance, authentication performance and data loss robustness because they are three major concerns in this paper.  Besides the assumptions in section II, we also assume 256-QAM modulation in all links. The advantage of using QAM is the ability to carry more bits of information per symbol, which provides high throughput.
%---------------------------------------------------------------------
\subsection{Secure transmission performance}
\begin{figure}[!ht]
	\centering
	\includegraphics[width=8.5cm]{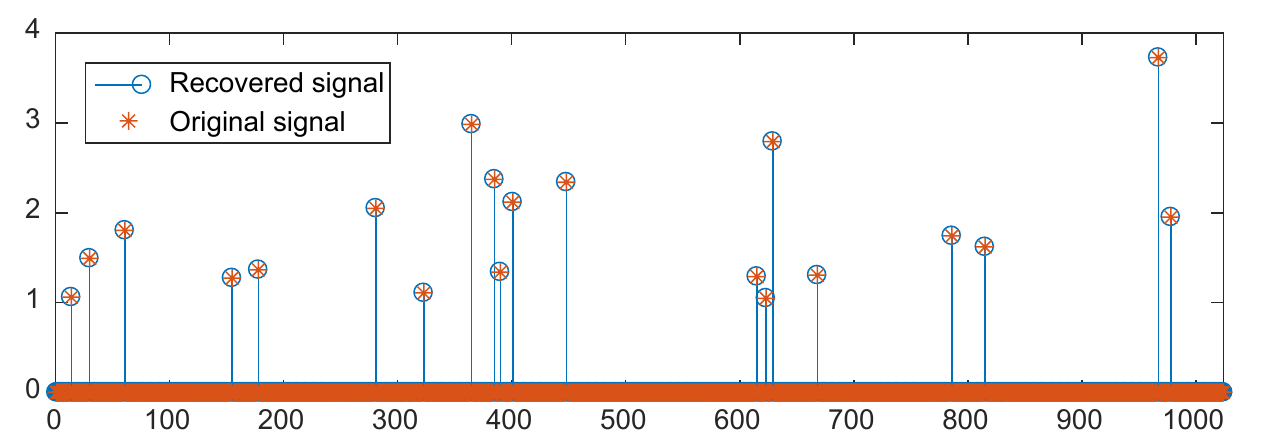}
	\caption{The original signal and the recovered signal.}
	\label{fig6}
\end{figure}

\begin{figure}[!ht]
	\centering
	\includegraphics[width=8.5cm]{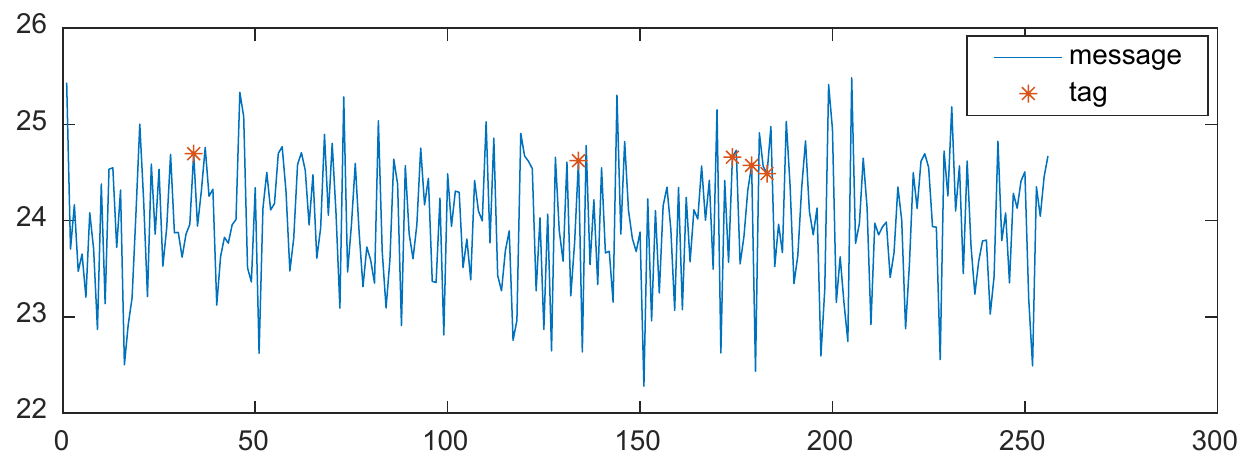}
	\caption{The transmitting message which contain both message information and tag information.}
	\label{fig7}
\end{figure}

 Quantitative evaluation is carried out using Monte Carlo simulation of simulated sparse signals. This is reasonable because a sparse decomposition block can be used to pre-process the original data. Sparse signals with different sparsity level are generated for evaluation. The signal length $n = 1024$, the sampling number $m = 256$ . Under such settings, an additional benefit is that the transmission burden reduced to 40\% comparing to transmitting the original signal, which will save power for the transmitter. Fig. \ref{fig6} shows such a signal which has small part of random non-zeros values. The corresponding transmitting message is shown in Fig. \ref{fig7}, where the tags randomly hide in the message. And the tag values are in the same amplitude level as message. Note that the tag number is variable making it more robust to crack. Under MIMO Rayleigh fading channel and MRC receiver as well as QAM modulation, the recovered signal using OMP reconstruction method is shown as star stems in Fig. \ref{fig6}. This indicates a perfect recovery, i.e., good transmission quality. To define perfect recovery, we use RMSE between the original signal and the recovered signal and let it below certain threshold $\tau_3$,
\begin{equation}
RMSE = \sqrt {\frac{{{{\left\| {{\bf{\hat x}} - {\bf{x}}} \right\|}_2}}}{n}}  < {\tau _3}
\label{eq15}
\end{equation}

Signal-to-noise-ratio (SNR) is one of the core issues in physical layer according to Shannon equation, which is directly related to transmission quality. As discussed in section III.D, some filtering process can be added to the demodulated signal to get rid of wrong transmission symbols which is caused by low SNR. The receiver can still perfectly recover under such case. So, the proposed scheme has certain degree of error correction ability from this perspective. To investigate the influence on of SNR the recovery probability, Monte Carlo simulation is carried out to obtain the perfect recovery probability, which is the ratio of perfect recovery number and total simulation number. All simulations are conducted under same Rayleigh fading parameter $\Omega$. Complex Gaussian noise are introduced and assuming no data loss temporarily.
\begin{figure}[!ht]
	\centering
	\includegraphics[width=8.5cm]{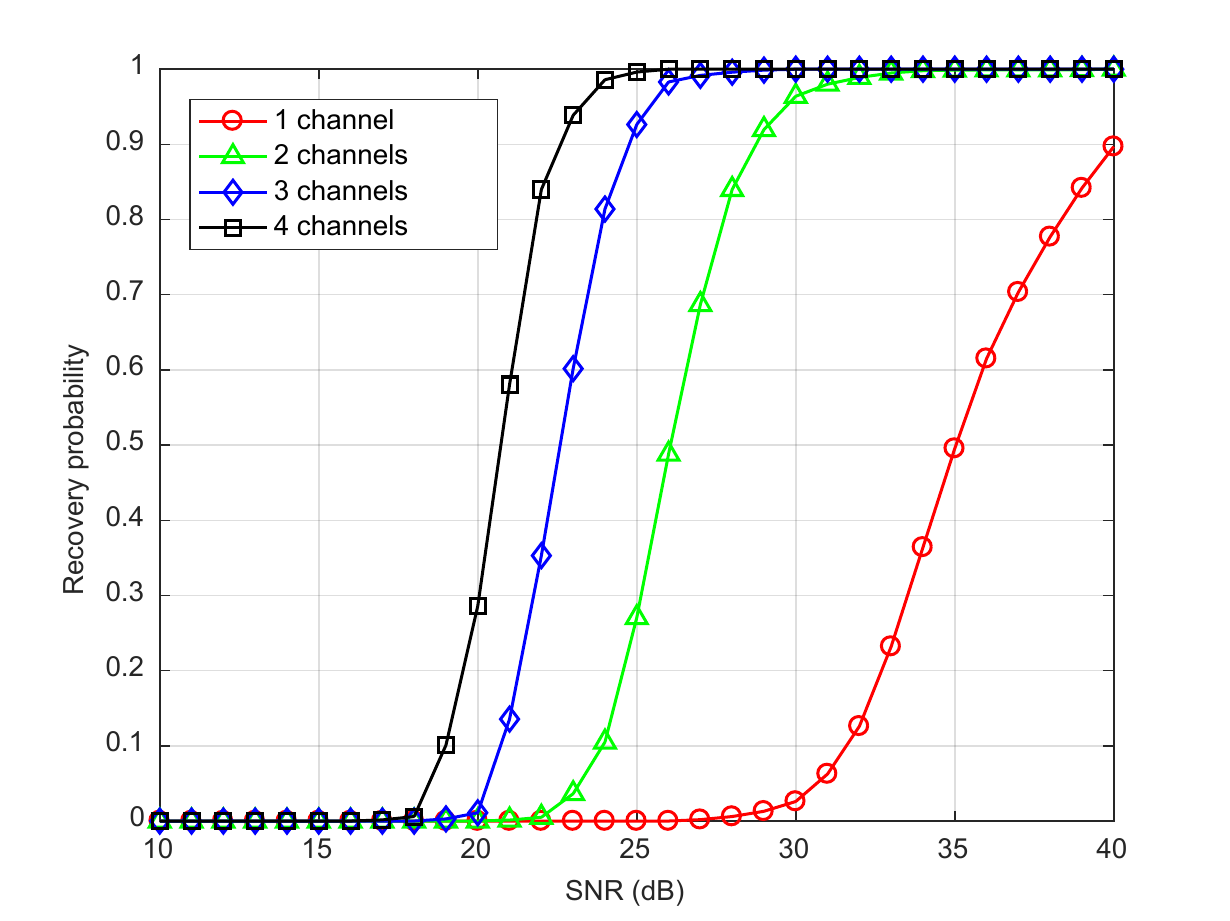}
	\caption{The influence of SNR on recovery probability under different number of MIMO channels. Sparsity level is 2\%.}
	\label{fig8}
\end{figure}
\begin{figure}[!ht]
	\centering
	\includegraphics[width=8.5cm]{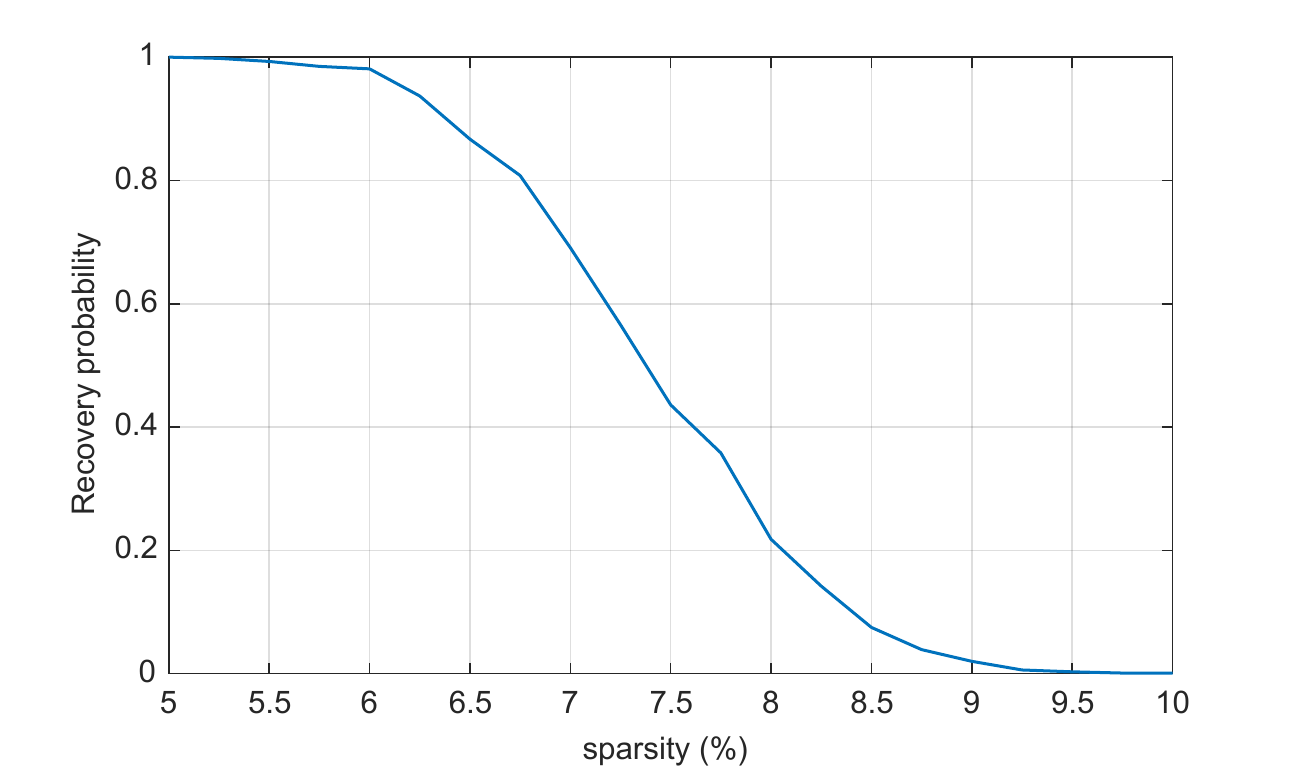}
	\caption{The influence of sparsity level on the recovery probability under perfect transmission.}
	\label{fig9}
\end{figure}

The influence of SNR on the recovery probability under different number of MIMO channels are shown in Fig. \ref{fig8}. The recovery probability increases sharply after certain threshold. The corresponding theory is (\ref{eq6}) which indicates that perfect recovery is guaranteed if the sampling number is large enough. While we set sampling number as constant value here, the final received measurement results will reduce due to throwing significant wrong symbols. More MIMO channels have better recovery probability because SNR is improved under more MIMO channel using MRC. Besides the influence of SNR on the final received measurement results, the sparsity level influences the recovery probability as well. To eliminate the interference of SNR, we assume perfect transmission which means all symbols are correctly transmitted. It is obvious that the recovery probability decreases with the increase of sparsity. This means that a sparser representation is helpful to the recovery probability.
%---------------------------------------------------------------
\subsection{Authentication performance}
\begin{figure}[!ht]
	\centering
	\includegraphics[width=8.5cm]{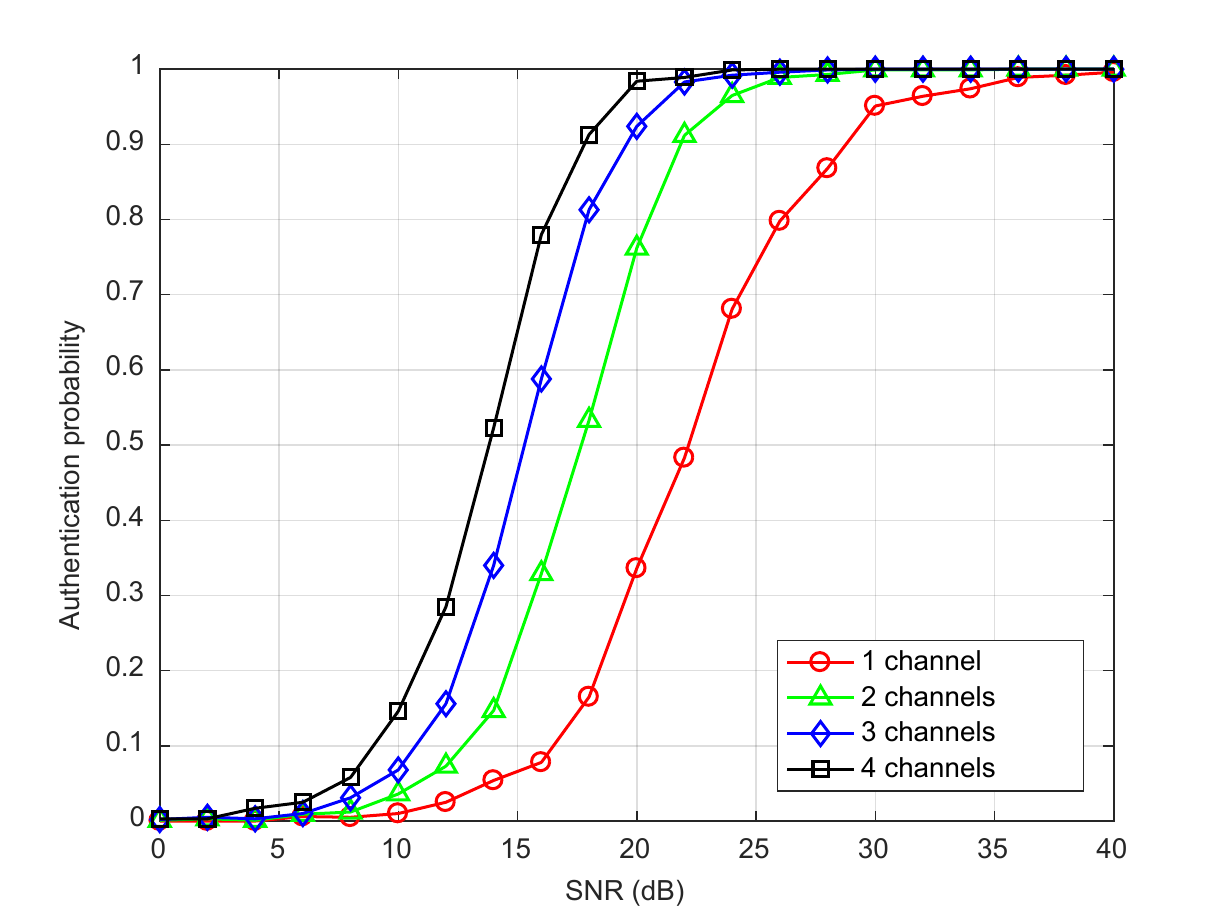}
	\caption{The influence of SNR on authentication probability under different MIMO channels. Sparsity level is 2\%.}
	\label{fig10}
\end{figure}

During the authentication process, the receiver compares the extracted tag in the received signal and the generated tag. To evaluate the authentication performance, the authentication probability \cite{RN529} is adopted as a metric. SNR will influence the authentication probability because transmission error in tag harms the authentication process. We assume no data loss here again and the sparsity level is set as 2\%. $\tau_2$  is set as 0.6 which means 40\% of tag errors are allowed. The influence of SNR on authentication probability under different MIMO channels are shown in Fig. \ref{fig10}. Compared to Fig. \ref{fig8}, the authentication probability under same SNR and MIMO channels are higher than that of recovery probability. This indicates that the authentication probabilities are less sensitive to transmission errors. Because the chance that an error symbol appears in tags are lower than that of data in the message.
%\vspace{-0.5cm}
%------------------------------------------------
\subsection{Data loss robustness}
\begin{figure}[!ht]
	\centering
	\includegraphics[width=8.5cm]{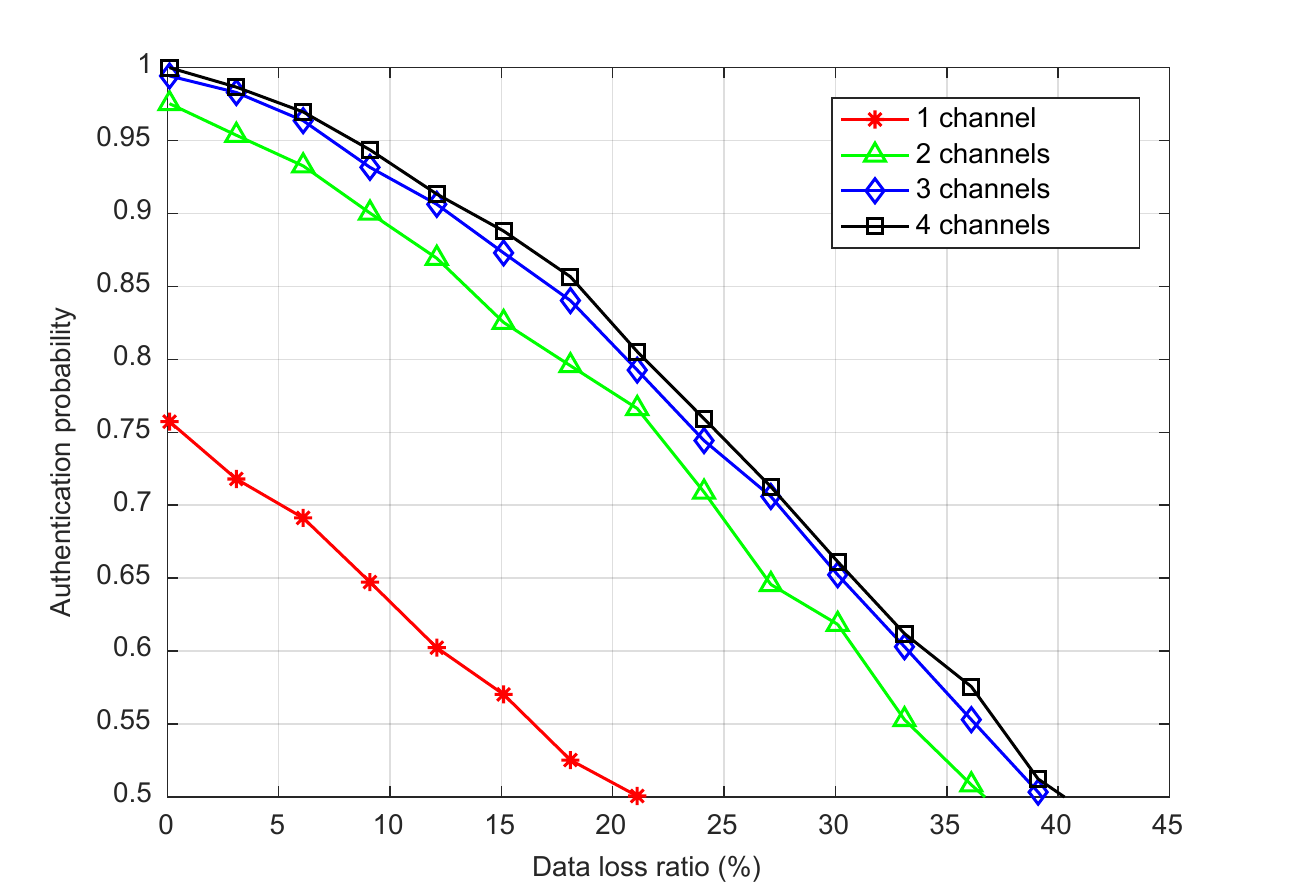}
	%\vspace{-0.3cm}
	\caption{The influence of data loss ratio on authentication probability under different MIMO channels. SNR = 40dB.}
	\label{fig11}
\end{figure}

\begin{figure}[!ht]
	%\vspace{-0.5cm}
	\centering
	\includegraphics[width=8.5cm]{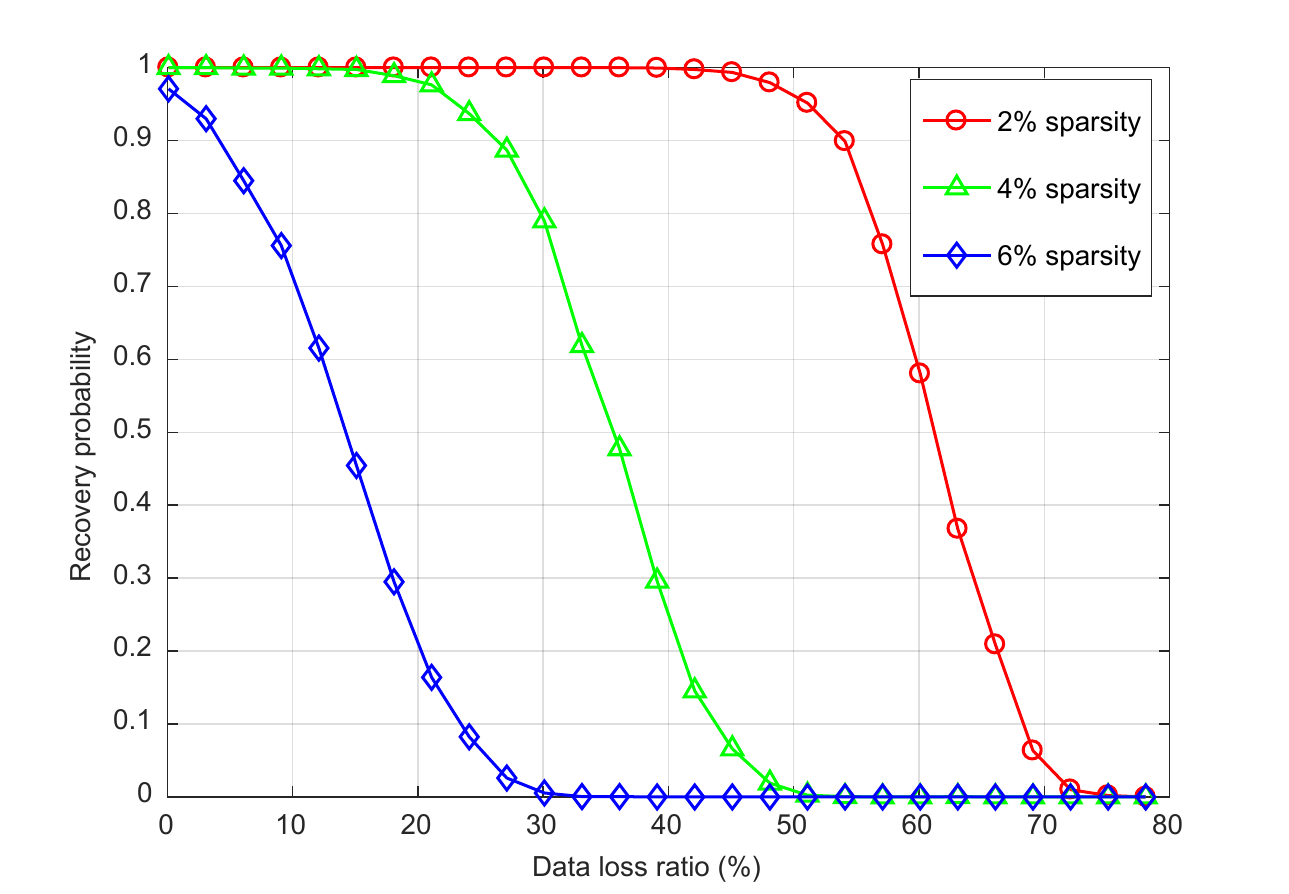}
	%\vspace{-0.3cm}
	\caption{The influence of data loss ratio on recovery probability under different sparsity level with 4 MIMO channels. SNR = 40dB.}
	\label{fig12}
\end{figure}

This subsection investigates the influence of data loss ratio on both authentication performance and secure transmission performance. Data loss ratio is the ratio between the number of lost data and total sending number. Different percentage of data loss ratio in the whole transmitting message are simulated under different MIMO channel quantity. The influence on authentication probability is shown in Fig. \ref{fig11} and Fig. \ref{fig12} shows the influence on recovery probability under different sparsity level. Fig. \ref{fig11} indicates that more than half of authentication is successful even suffer from 40\% of data loss in the case of 4 MIMO channels. These curves are less steep and the one MIMO channel curve is far away from other curves, both observations indicate that the authentication probability is more sensitive to SNR than data loss. As for the influence on recovery probability, the data loss ratio which starts decreasing almost proportional to the sparsity level, the recovery probability decreases sharply after this ratio. These phenomena coincide with (\ref{eq6}). Amazingly, perfect recovery is still guaranteed under 40\% of data loss for 2\% of sparsity level, which means that the proposed scheme is robust to data loss.
%=====================================
\section{Conclusions}
This paper proposed a compressed sensing-based framework which achieves encryption and authentication concurrently with a physical layer design for the first time. The framework offers a well-integrated security solution that efficiently safeguards the confidential/privacy/robust data communication in cyber-physical systems of Internet of Things (IoT). It is beneficial for safety-critical applications like telemedicine and structural health monitoring. This framework contains 1) a novel 2D key (measurement matrix) generation algorithm which is much more time-efficient than state-of-the-art methods. It also eliminates pre-shared key and complex key management; 2) a novel authentication information generation process which eliminates complex computation; 3) a novel signal structure which concurrently transmit data and authentication information; 4) a novel authentication scheme which is robust to data loss and transmission error; 5) a data recovery scheme which is able to correct transmission error and data loss caused in the physical layer. Although the first three points are based on channel gain, it is easy to transfer to other channel state information attributes. Under this framework, the data access latency is reduced due to encryption and authentication fusion. Continuous authentication can be achieved as well because authentication information and data go parallelly.
%\vspace{-0.5cm}

%\begin{acknowledgements}
%If you'd like to thank anyone, place your comments here
%and remove the percent signs.
%\end{acknowledgements}

% Authors must disclose all relationships or interests that 
% could have direct or potential influence or impart bias on 
% the work: 
%
% \section*{Conflict of interest}
%
% The authors declare that they have no conflict of interest.

% BibTeX users please use one of
%\bibliographystyle{spbasic}      % basic style, author-year citations
%\bibliographystyle{spmpsci}      % mathematics and physical sciences
%\bibliographystyle{spphys}       % APS-like style for physics
%\bibliography{}   % name your BibTeX data base

% Non-BibTeX users please use
%\begin{thebibliography}{}
%
% and use \bibitem to create references. Consult the Instructions
% for authors for reference list style.
%
%\bibitem{RefJ}
% Format for Journal Reference
%Author, Article title, Journal, Volume, page numbers (year)
% Format for books
%\bibitem{RefB}
%Author, Book title, page numbers. Publisher, place (year)
% etc
%\end{thebibliography}

\bibliographystyle{spmpsci}
\bibliography{bare_jrnl_comsoc}

\begin{thebibliography}{10}
\providecommand{\url}[1]{{#1}}
\providecommand{\urlprefix}{URL }
\expandafter\ifx\csname urlstyle\endcsname\relax
  \providecommand{\doi}[1]{DOI~\discretionary{}{}{}#1}\else
  \providecommand{\doi}{DOI~\discretionary{}{}{}\begingroup
  \urlstyle{rm}\Url}\fi

\bibitem{RN533}
Baracca, P., Laurenti, N., Tomasin, S.: Physical layer authentication over
  \protect{MIMO} fading wiretap channels.
\newblock IEEE Trans. Wireless Commun. \textbf{11}(7), 2564--2573 (2012).
\newblock \doi{10.1109/TWC.2012.051512.111481}

\bibitem{RN35}
Barcelo-Llado, J.E., Morell, A., Seco-Granados, G.: Amplify-and-forward
  compressed sensing as an energy-efficient solution in wireless sensor
  networks.
\newblock IEEE Sensors J. \textbf{14}(5), 1710--1719 (2014)

\bibitem{RN88}
Cai, T.T., Wang, L.: Orthogonal matching pursuit for sparse signal recovery
  with noise.
\newblock IEEE Trans. Inf. Theory \textbf{57}(7), 4680--4688 (2011)

\bibitem{RN766}
Cambareri, V., Mangia, M., Pareschi, F., Rovatti, R., Setti, G.: Low-complexity
  multiclass encryption by compressed sensing.
\newblock IEEE Trans. Signal Process. \textbf{63}(9), 2183--2195 (2015).
\newblock \doi{10.1109/TSP.2015.2407315}

\bibitem{RN4}
Cand\'{e}s, E.J., Romberg, J., Tao, T.: Robust uncertainty principles: Exact
  signal reconstruction from highly incomplete frequency information.
\newblock IEEE Trans. Inf. Theory \textbf{52}(2), 489--509 (2006)

\bibitem{RN272}
Cand\'{e}s, E.J., Romberg, J.K., Tao, T.: Stable signal recovery from
  incomplete and inaccurate measurements.
\newblock Comm. Pure Appl. Math. \textbf{59}(8), 1207--1223 (2006)

\bibitem{RN559}
Ciaramella, A., Giunta, G.: Packet loss recovery in audio multimedia streaming
  by using compressive sensing.
\newblock IET Commun. \textbf{10}(4), 387--392 (2016).
\newblock \doi{10.1049/iet-com.2014.0995}

\bibitem{RN457}
Dautov, R., Tsouri, G.R.: Securing while sampling in wireless body area
  networks with application to electrocardiography.
\newblock IEEE J. Biomed. Health Inform. \textbf{20}(1), 135--142 (2016).
\newblock \doi{10.1109/JBHI.2014.2366125}

\bibitem{RN549}
Dean, T.R., Goldsmith, A.J.: Physical-layer cryptography through massive mimo.
\newblock IEEE Trans. Inf Theory \textbf{63}(8), 5419--5436 (2017).
\newblock \doi{10.1109/TIT.2017.2715187}

\bibitem{RN310}
Donoho, D.L., Huo, X.: Uncertainty principles and ideal atomic decomposition.
\newblock IEEE Trans. Inf Theory \textbf{47}(7), 2845--2862 (2001)

\bibitem{RN526}
Ferrante, A., Laurenti, N., Masiero, C., et~al.: On the error region for
  channel estimation-based physical layer authentication over
  \protect{R}ayleigh fading.
\newblock IEEE Trans. Inf. Forensics Security \textbf{10}(5), 941--952 (2015)

\bibitem{RN538}
Hao, P., Wang, X., Behnad, A.: Relay authentication by exploiting \protect{I/Q}
  imbalance in amplify-and-forward system.
\newblock In: Global Communications Conference (GLOBECOM), 2014 IEEE, pp.
  613--618. IEEE

\bibitem{RN541}
Hooshmand, R., Aref, M.R.: Efficient polar code-based physical layer encryption
  scheme.
\newblock IEEE Wireless Commun. Lett. \textbf{6}(6), 710--713 (2017)

\bibitem{RN540}
Hou, W., Wang, X., Chouinard, J.Y., et~al.: Physical layer authentication for
  mobile systems with time-varying carrier frequency offsets.
\newblock IEEE Trans. Commun. \textbf{62}(5), 1658--1667 (2014)

\bibitem{RN556}
Kailkhura, B., Liu, S., Wimalajeewa, T., Varshney, P.K.: Measurement matrix
  design for compressed detection with secrecy guarantees.
\newblock IEEE Wireless Commun. Lett. \textbf{5}(4), 420--423 (2016).
\newblock \doi{10.1109/LWC.2016.2574853}

\bibitem{RN555}
Kailkhura, B., Wimalajeewa, T., Varshney, P.K.: Collaborative compressive
  detection with physical layer secrecy constraints.
\newblock IEEE Trans. Signal Process. \textbf{65}(4), 1013--1025 (2017).
\newblock \doi{10.1109/TSP.2016.2630029}

\bibitem{RN546}
Kanter, G.S., Reilly, D., Smith, N.: Practical physical-layer encryption: The
  marriage of optical noise with traditional cryptography.
\newblock IEEE Commun. Mag. \textbf{47}(11), 74--81 (2009).
\newblock \doi{10.1109/MCOM.2009.5307469}

\bibitem{RN561}
Khansefid, A., Minn, H.: Achievable downlink rates of mrc and zf precoders in
  massive \protect{MIMO} with uplink and downlink pilot contamination.
\newblock IEEE Trans. Commun. \textbf{63}(12), 4849--4864 (2015).
\newblock \doi{10.1109/TCOMM.2015.2482965}

\bibitem{RN523}
Lin, J., Yu, W., Zhang, N., Yang, X., et~al.: A survey on internet of things:
  architecture, enabling technologies, security and privacy, and applications.
\newblock IEEE Internet Things J. \textbf{4}(57)

\bibitem{RN552}
Liu, C., Yang, N., Yuan, J., Malaney, R.: Location-based secure transmission
  for wiretap channels.
\newblock IEEE J. Sel. Areas Commun. \textbf{33}(7), 1458--1470 (2015).
\newblock \doi{10.1109/JSAC.2015.2430211}

\bibitem{RN530}
Liu, J., Wang, X.: Physical layer authentication enhancement using
  two-dimensional channel quantization.
\newblock IEEE Trans. Wireless Commun. \textbf{15}(6), 4171--4182 (2016).
\newblock \doi{10.1109/TWC.2016.2535442}

\bibitem{RN558}
Mayiami, M.R., Seyfe, B., Bafghi, H.G.: Perfect secrecy via compressed sensing.
\newblock In: 2013 Iran Workshop on Communication and Information Theory, pp.
  1--5.
\newblock \doi{10.1109/IWCIT.2013.6555751}

\bibitem{RN550}
Mostafa, A., Lampe, L.: Physical-layer security for \protect{MISO} visible
  light communication channels.
\newblock IEEE J. Sel. Areas Commun. \textbf{33}(9), 1806--1818 (2015)

\bibitem{RN537}
Mukherjee, A., Fakoorian, S.A.A., Huang, J., Swindlehurst, A.L.: Principles of
  physical layer security in multiuser wireless networks: A survey.
\newblock IEEE Commun. Surveys Tuts. \textbf{16}(3), 1550--1573 (2014).
\newblock \doi{10.1109/SURV.2014.012314.00178}

\bibitem{RN539}
Polak, A.C., Dolatshahi, S., Goeckel, D.L.: Identifying wireless users via
  transmitter imperfections.
\newblock IEEE J. Sel. Areas Commun. \textbf{29}(7), 1469--1479 (2011)

\bibitem{RN25}
Qaisar, S., Bilal, R.M., Iqbal, W., Naureen, M., Lee, S.: Compressive sensing:
  From theory to applications, a survey.
\newblock J. Commun. Netw. \textbf{15}(5), 443--456 (2013)

\bibitem{RN543}
Sakai, M., Lin, H., Yamashita, K.: Intrinsic interference based physical layer
  encryption for \protect{OFDM/OQAM}.
\newblock IEEE Commun. Lett. \textbf{21}(5), 1059--1062 (2017).
\newblock \doi{10.1109/LCOMM.2017.2654442}

\bibitem{RN534}
Shan, D., Zeng, K., Xiang, W., Richardson, P., Dong, Y.: \protect{PHY-CRAM}:
  Physical layer challenge-response authentication mechanism for wireless
  networks.
\newblock IEEE J. Sel. Areas Commun. \textbf{31}(9), 1817--1827 (2013).
\newblock \doi{10.1109/JSAC.2013.130914}

\bibitem{RN535}
Shi, L., Li, M., Yu, S., Yuan, J.: \protect{BANA}: Body area network
  authentication exploiting channel characteristics.
\newblock IEEE J. Sel. Areas Commun. \textbf{31}(9), 1803--1816 (2013).
\newblock \doi{10.1109/JSAC.2013.130913}

\bibitem{RN460}
Tang, C., Tian, G.Y., Li, K., Sutthaweekul, R., Wu, J.: Smart compressed
  sensing for online evaluation of cfrp structure integrity.
\newblock IEEE Trans. Ind. Electron. \textbf{64}(12), 9608--9617 (2017).
\newblock \doi{10.1109/tie.2017.2698406}

\bibitem{RN564}
Tokognon, C.A., Gao, B., Tian, G.Y., Yan, Y.: Structural health monitoring
  framework based on internet of things: A survey.
\newblock IEEE Internet Things J. \textbf{4}(3), 619--635 (2017).
\newblock \doi{10.1109/JIOT.2017.2664072}

\bibitem{RN18}
Tropp, J.A., Gilbert, A.C.: Signal recovery from random measurements via
  orthogonal matching pursuit.
\newblock IEEE Trans. Inf. Theory \textbf{53}(12), 4655--4666 (2007)

\bibitem{RN536}
Tugnait, J.K.: Wireless user authentication via comparison of power spectral
  densities.
\newblock IEEE J. Sel. Areas Commun. \textbf{31}(9), 1791--1802 (2013).
\newblock \doi{10.1109/JSAC.2013.130912}

\bibitem{RN525}
Verma, G., Yu, P., Sadler, B.M.: Physical layer authentication via fingerprint
  embedding using software-defined radios.
\newblock IEEE Access \textbf{3}, 81--88 (2015)

\bibitem{RN522}
Wang, C.X., Haider, F., Gao, X., You, X.H., et~al.: Cellular architecture and
  key technologies for \protect{5G} wireless communication networks.
\newblock IEEE Commun. Mag. \textbf{52}(2), 122--130 (2014)

\bibitem{RN514}
Wang, N., Jiang, T., Li, W., Lv, S.: Physical-layer security in internet of
  things based on compressed sensing and frequency selection.
\newblock IET Commun. \textbf{11}(9), 1431--1437 (2017).
\newblock \doi{10.1049/iet-com.2016.1088}

\bibitem{RN529}
Wang, X., Hao, P., Hanzo, L.: Physical-layer authentication for wireless
  security enhancement: current challenges and future developments.
\newblock IEEE Commun. Mag. \textbf{54}(6), 152--158 (2016).
\newblock \doi{10.1109/MCOM.2016.7498103}

\bibitem{RN553}
Wilhelm, M., Martinovic, I., Schmitt, J.B.: Secure key generation in sensor
  networks based on frequency-selective channels.
\newblock IEEE J. Sel. Areas Commun. \textbf{31}(9), 1779--1790 (2013).
\newblock \doi{10.1109/JSAC.2013.130911}

\bibitem{RN551}
Wu, C.Y., Lan, P.C., Yeh, P.C., Lee, C.H., Cheng, C.M.: Practical physical
  layer security schemes for \protect{MIMO-OFDM} systems using precoding matrix
  indices.
\newblock IEEE J. Sel. Areas Commun. \textbf{31}(9), 1687--1700 (2013)

\bibitem{RN527}
Wu, X.: Embedded physical-layer authentication in cognitive radio requires
  efficient low-rate channel coding schemes.
\newblock IET Commun. \textbf{11}(3), 400--404 (2017).
\newblock \doi{10.1049/iet-com.2016.0812}

\bibitem{RN532}
Wu, X., Yang, Z.: Physical-layer authentication for multi-carrier transmission.
\newblock IEEE Commun. Lett. \textbf{19}(1), 74--77 (2015).
\newblock \doi{10.1109/LCOMM.2014.2375191}

\bibitem{RN528}
Wu, X., Yang, Z., Ling, C., Xia, X.G.: Artificial-noise-aided physical layer
  phase challenge-response authentication for practical \protect{OFDM}
  transmission.
\newblock IEEE Trans. Wireless Commun. \textbf{15}(10), 6611--6625 (2016)

\bibitem{RN524}
Yang, Y., Wu, L., Yin, G., Li, L., Zhao, H.: A survey on security and privacy
  issues in internet-of-things.
\newblock IEEE Internet Things J. \textbf{4}(5), 1250--1258 (2017).
\newblock \doi{10.1109/JIOT.2017.2694844}

\bibitem{RN767}
Yang, Z., Zhang, C., Xie, L.: Robustly stable signal recovery in compressed
  sensing with structured matrix perturbation.
\newblock IEEE Trans. Signal Process. \textbf{60}(9), 4658--4671 (2012).
\newblock \doi{10.1109/TSP.2012.2201152}

\bibitem{RN531}
Yu, P.L., Verma, G., Sadler, B.M.: Wireless physical layer authentication via
  fingerprint embedding.
\newblock IEEE Commun. Mag. \textbf{53}(6), 48--53 (2015).
\newblock \doi{10.1109/MCOM.2015.7120016}

\bibitem{RN544}
Zhang, J., Marshall, A., Woods, R., Duong, T.Q.: Design of an \protect{OFDM}
  physical layer encryption scheme.
\newblock IEEE Trans. Veh. Technol. \textbf{66}(3), 2114--2127 (2017).
\newblock \doi{10.1109/TVT.2016.2571264}

\bibitem{RN38}
Zou, Z., Bao, Y., Li, H., Spencer, B.F., Ou, J.: Embedding compressive
  sensing-based data loss recovery algorithm into wireless smart sensors for
  structural health monitoring.
\newblock IEEE Sensors J. \textbf{15}(2), 797--808 (2015)

\end{thebibliography}

\end{document}